\documentclass[prb,superscriptaddress,showpacs,twocolumn]{revtex4}

\usepackage{graphicx}

\usepackage{amssymb,amsmath}
\usepackage{subfigure}
\usepackage[dvips]{color}

\newcommand{\eref}[1]{Eq.~(\ref{#1})}
\newcommand{\fref}[1]{Figure~\ref{#1}}

\newcommand{\im}{%
           \imath}
\newcommand{\bra}[1]{\ensuremath{\langle #1|}}
\newcommand{\ket}[1]{\ensuremath{|#1\rangle}}

\newcommand{\bracket}[2]{\langle #1|#2\rangle}

\newcommand{\svek}{%
        \mathbf}

\newcommand{\vek}[1]{%
        \hbox{\textbf #1}}

\newcommand{\pr}{%
        ^\prime}

\def\XXint#1#2#3{{\setbox0=\hbox{$#1{#2#3}{\int}$}
\vcenter{\hbox{$#2#3$}}\kern-.5\wd0}}

\newcommand{\oo}[1]{\frac{1}{#1}}

\begin{document}

\author{ Jan M. Tomczak}
\affiliation{Research Institute for Computational Sciences, AIST, Tsukuba, 305-8568 Japan}
\affiliation{Japan Science and Technology Agency, CREST, Kawaguchi, Japan}
\author{ T. Miyake}
\affiliation{Research Institute for Computational Sciences, AIST, Tsukuba, 305-8568 Japan}
\affiliation{Japan Science and Technology Agency, CREST, Kawaguchi, Japan}
\author{ R. Sakuma}
\affiliation{Graduate School of Advanced Integration Science, Chiba University, Chiba, 263-8522 Japan}
\affiliation{Japan Science and Technology Agency, CREST, Kawaguchi, Japan}
\author{ F. Aryasetiawan}
\affiliation{Graduate School of Advanced Integration Science, Chiba University, Chiba, 263-8522 Japan}
\affiliation{Research Institute for Computational Sciences, AIST, Tsukuba, 305-8568 Japan}
\affiliation{Japan Science and Technology Agency, CREST, Kawaguchi, Japan}

\title{Effective Coulomb interactions in solids under pressure}

\begin{abstract}
Correlated materials are extremely sensitive to external stimuli, such as temperature or pressure.
Describing the electronic properties of such systems often requires applying many-body techniques to
effective low energy problems in the spirit of the Hubbard model, or extensions thereof.
While the effect of pressure on structures and bands 
has been investigated extensively within density-functional based methods, 
the pressure dependence of electron-electron interactions has so far received little attention.
As a step toward {\it ab initio }pressure studies for
realistic systems within a setup of maximally localized
Wannier functions and the constrained random phase approximation, we examine in this paper the
paradigmatic pressure dependence of Coulomb interactions.
While compression commonly causes the ``extension'' of  Wannier functions, and thus transfer elements, to grow,
we find the -- seemingly counter-intuitive -- tendency that the bare Coulomb interaction {\it increases} under compression as well.
We reconcile these behaviors by appealing to a semi-analytical tight-binding model. We moreover argue that, for this model,
the requirement of maximal Wannier localization is equivalent to maximizing the Coulomb interaction matrix elements.
We then apply the above first principles techniques to fcc hydrogen under pressure.
While we find our comprehension of the bare Coulomb interaction confirmed, the induced changes in screening
strengths lead to an effective one band model with a Hubbard interaction that is non-monotonous under pressure.
\end{abstract}

\maketitle

\section{Introduction}
The panoply of structural, orbital and spin degrees of freedom and the joint presence of
important electronic Coulomb interactions
 cause correlated materials~\cite{imada} 
to be the realm neither of band-theory nor of model many-body physics each on their own.
Therefore, in recent years, much ingenuity was invested into finding ways to merge the ``realism'' of the former
with the accurate description of correlation effects of the latter.

With the exception of the GW approximation~\cite{hedin,ferdi_gw,RevModPhys.74.601} to Hedin's equations~\cite{hedin}, the electronic structure approach 
GW+DMFT~\cite{PhysRevLett.90.086402} that combines GW with the dynamical mean-field theory (DMFT)~\cite{bible}, and a recent proposal for a self-energy downfolding~\cite{ferdi_down},
this combining commonly amounts 
to extracting a low-energy one-particle Hamiltonian from density-functional-theory-based methods, such as the local-density approximation (LDA),~\cite{RevModPhys.71.1253,RevModPhys.61.689}
supplementing it with interaction terms [e.g.\ of the Hubbard-Hund (U and J) type] and to solving the resulting ``realistic model'' with a chosen many-body technique.

Hence, in this approach there are two intertwined preparative tasks that generate the many-body problem.
The deducing of the low-energy one-particle part can be achieved e.g.\ by a tight binding fit of relevant bands, the 
downfolding~\cite{loewdin,PhysRevB.34.2439} procedure within e.g.\ muffin-tin-based methods,~\cite{PhysRevB.34.2439,lmto,nmto} or by the
generation of (maximally localized~\cite{PhysRevB.56.12847,PhysRevB.65.035109})
Wannier functions.\cite{RevModPhys.34.645}

The interaction parameters of realistic many-body models, in turn, are often chosen rather empirically than from  a solid first principles basis, a fact that has caused many objections in the past. In particular, when tracking properties as a function of an external parameter -- pressure, in our case -- the evolution of the interaction has mostly been  discarded.

This and the quest for going beyond mere qualitative results towards, eventually, the quantitative design of materials, 
 highlights the need for accurate ways to determine all ingredients of realistic models in  an {\it ab initio} 
fashion.
Nowadays, the most popular methods for the computation of interaction matrix elements are
the constrained LDA
technique~\cite{constrainedLDA},  and the constrained random phase approximation (cRPA)~\cite{PhysRevB.70.195104}.

A recent, and promising approach is the use of Wannier functions within the cRPA setup~\cite{miyake:085122}, which allows
for a deducing of the one-particle and two-particle parts of the Hamiltonian on the same footing.
Moreover, working in a localized Wannier type of basis is often a requirement of many-body approaches such as the DMFT~\cite{bible}.
As to the interaction matrix elements, the cRPA technique allows for a precise elimination of
the screening channels of the chosen orbital subspace that constitutes the effective model~\cite{PhysRevB.70.195104}.

While these techniques have already been applied for the setting up of many-body models of some
complex materials~\cite{miyake:085122,JPSJS.77.093711,JPSJS.77SC.99}, a basic understanding of the 
effects of pressure on the Coulomb interactions within a Wannier setup is lacking.
This is the aim of the current work.

In a first part, we investigate a semi-analytical tight binding model of a one-band solid in one dimension, track transfer matrix elements, the bare (i.e.\ unscreened) Coulomb interaction, and the spread of the maximally localized Wannier orbitals as a function of lattice spacing.
Being able to access the decomposition of the maximally localized Wannier functions onto the tight-binding basis will allow to understand
the surprising finding that under pressure the Coulomb interaction matrix element {\it augments}, while, at the same time,
transfer elements describing the delocalization of electrons {\it grow} as well.
As a more realistic example, we, second, apply the fully {\it ab initio} approach of the cRPA within maximally localized Wannier functions~\cite{miyake:085122} to fcc hydrogen, which is found to exhibit the explained generic behavior of the bare Coulomb interaction matrix elements. However, the partially screened Coulomb interaction -- the Hubbard $U$ of an effective low energy model for the half-filled 1s orbital -- actually shows a non-monotonous trend -- a consequence of two opposing effects in screening processes.
\section{Method}
The method of using the Wannier orbital construction in conjunction with the cRPA technique has been presented in detail in 
Ref.~\cite{miyake:085122}. 
While not fully reviewing the approach, we will discuss some issues relevant for the understanding of our results and
introduce some notation.

For the one-particle band-structure, a density functional calculation is performed. For the realistic case of fcc hydrogen, we will
employ the LDA~\cite{RevModPhys.61.689} in the full-potential linear muffin-tin orbital (FP-LMTO)~\cite{fplmto} realization. For obtaining the random phase approximation (RPA) polarizations we employ
the code of Ref.~\cite{fpgw} with the maximally localized Wannier extension of Ref.~\cite{miyake:085122},
and construct an effective system for the isolated 1s orbital. That is, we introduce the sub-Hilbert space 
$\mathcal{H}^{\hbox{\tiny eff}}=span\left\{{\ket{\psi^{\hbox{\tiny KS}\phantom{*}}_{\svek{k}1s}}}\right\}$,
spanned by the 1s Kohn-Sham wave function.
Since the aim is to work within a localized basis, the
extraction of the low-energy part is done by
a construction
of Wannier functions~\cite{RevModPhys.34.645} for $\mathcal{H}^{\hbox{\tiny eff}}$, as described in  Refs.~\cite{PhysRevB.56.12847,PhysRevB.65.035109,miyake:085122}.
The corresponding effective interactions are then computed within the constrained RPA~\cite{PhysRevB.70.195104} formalism.
This amounts to screening the
matrix elements of the bare Coulomb interaction $v(\svek{r},\svek{r}\pr)=1/\left| \svek{r}-\svek{r}\pr\right|$, which in the Wannier basis are given
by
\begin{eqnarray}
\label{eqV}
		&&V^{\alpha\beta\alpha\pr\beta\pr}_{\svek{R},\svek{R}\pr}=\\
		&&\quad\int d^3r d^3r\pr \chi^{\hbox{\tiny 
		W}*}_{\svek{R}\alpha}(\svek{r}) \chi^{\hbox{\tiny W}}_{\svek{R}\beta}(\svek{r})\frac{1}{\left|\vek{r}-\vek{r}\pr\right|}
	\chi^{\hbox{\tiny W}*}_{{\svek{R}\pr}\alpha\pr}(\svek{r}\pr) \chi^{\hbox{\tiny W}}_{\svek{R}\pr\beta\pr}(\svek{r}\pr)\nonumber
\end{eqnarray}
with a partial RPA polarization 
\begin{equation}
P_r=P-P_s	
	\label{Pr}
\end{equation}
where $P$ and $P_s$ are the polarizations of the full and the sub-Hilbert spaces, respectively. The latter $P_s$, when
using the Kohn-Sham orbitals, can be expressed as 
\begin{eqnarray}
	\label{pol}
	&&P_s(\vek{r},\vek{r}\pr,\omega)= \sum_{spin}\,\,
	\sum_{ \psi^{\hbox{\tiny KS}}_{\svek{k} \alpha}\in \mathcal{H}^{\hbox{\tiny eff}}    }^{occ}\,\,\sum_{ \psi^{\hbox{\tiny KS}}_{\svek{k}\pr \beta}\in \mathcal{H}^{\hbox{\tiny eff}}}^{unocc} \nonumber\\
	 &&\quad\left\{ \frac{1}{\omega-\epsilon_{\svek{k}\pr \beta}+\epsilon_{\svek{k} \alpha} +\im 0^+} - \frac{1}{\omega+\epsilon_{\svek{k}\pr \beta}-\epsilon_{\svek{k} \alpha} - \im 0^+}    \right\}\nonumber\\
	 \nonumber\\
	 	&&\quad\times\quad\psi^{\hbox{\tiny KS}*}_{\svek{k}\alpha}(\vek{r})\psi^{\hbox{\tiny KS}\phantom{*}}_{\svek{k}\pr \beta}(\vek{r})\psi^{\hbox{\tiny KS}*}_{\svek{k}\pr \beta}(\vek{r}\pr)\psi^{\hbox{\tiny KS}\phantom{*}}_{\svek{k}\alpha}(\vek{r}\pr) 
	\end{eqnarray}
i.e.\ by transitions restricted to the effective sub-system, in our case $\mathcal{H}^{\hbox{\tiny eff}}=span\left\{{\ket{\psi^{\hbox{\tiny KS}\phantom{*}}_{\svek{k}1s}}}\right\}$.
Within this notation, the strengths of screening channels are influenced by two effects~: the matrix elements [the overlap integrals of wave functions that occur when calculating matrix elements of $P$, in analogy to \eref{eqV}], and the energy differences of the Kohn-Sham excitations, $\epsilon_{\svek{k}n}$, that appear in the denominators.
The virtue used for the constraining is the fact that the screening contributions are additive~\cite{PhysRevB.70.195104}. 
Indeed, when using the above decomposition $P=P_s+P_r$, 
the fully screened interaction $W$ (of the GW formalism~\cite{hedin,ferdi_gw,RevModPhys.74.601}) 
can be given in terms of the partially screened interaction for the effective model of the one-band orbital, $W_r={v}/({1-P_rv})$,   by
$W={W_r}/{(1-P_sW_r)}$~\cite{PhysRevB.70.195104}.
The Hubbard $U$ of the 1s sub-system is given by the 
on-site Wannier function matrix element of $W_r$.

The major observation in this context is that the construction of Wannier functions, and thus also the determination of interaction matrix elements are not unique~\cite{RevModPhys.34.645}. Indeed a unitary transformation
applied to the periodic part of the wave functions, while preserving the Bloch functions, changes the Wannier orbitals.
This gauge freedom can be used to choose the Wannier basis that is most suitable for the final purpose.
The aim of the low-energy system is to correct for local (Hubbard-type) interaction effects.
To this end, there exist at least two proposals on how to choose an optimal Wannier basis set~:
in the maximally localized Wannier approach~\cite{PhysRevB.56.12847,PhysRevB.65.035109} the
extension (``spread'') 
of the Wannier functions is minimized~:
denoting the Wannier states by kets, $\chi^{\hbox{\tiny W}}_{\svek{R}\alpha}(\vek{r})=\bracket{\vek{r}}{\alpha\vek{R}}$, 
this spread can be chosen as~\cite{PhysRevB.56.12847}
\begin{equation}
	\Omega=\left\langle \vek{r}^2\right\rangle_{\mathcal{H}^{\hbox{\tiny eff}}}=\!\!\!\sum_{\left\{\left. \alpha\,\right| \,\,
	\ket{\alpha\svek{0}}\in \mathcal{H}^{\hbox{\tiny eff}}\right\}}\!   
	\left[ \bra{\alpha\vek{0}}r^2\ket{\alpha\vek{0}} - \left|\bra{\alpha\vek{0}}\vek{r}\ket{\alpha\vek{0}}\right|^2 \right]
	\label{spread}
\end{equation}
The minimization of this $\Omega$ is of course only one of the possible options to fix the Wannier functions, yet a very natural one, since it can e.g.\ be shown~\cite{PhysRevB.26.4269,PhysRevB.56.12847} that, in one dimension,
the resulting Wannier functions are eigenfunctions of the subset projected position operator, an intuitive criterion for real space localization~\footnote{In higher dimensions, the non-commutation of the components of the position operator impedes this property~\cite{PhysRevB.26.4269,PhysRevB.56.12847}.}.

In the second scheme, the screened local Coulomb interaction
matrix element -- the Hubbard $U$, as given by the on-site component of $W_r$ from above -- is maximized~\cite{RevModPhys.35.457,miyake:085122}
to determine the Wannier functions.
For the case of SrVO$_3$, it has been shown~\cite{miyake:085122} that both approaches yield very similar results.
Indeed, appealing to e.g.\ the equation of the bare Coulomb interaction matrix element in the Wannier basis, \eref{eqV},
 it seems plausible that a greater localization of Wannier functions (smaller $\Omega$) results in an increased interaction matrix element $V$.
For the simple model in one dimension that we discuss in the following, we in fact motivate the equivalence of the maximally localized Wannier functions
and the basis in which the (bare) Coulomb interaction matrix element is maximal.

However, we stress that the intuitive correspondence between a stronger localization of Wannier functions, in the sense of the spread $\Omega$, 
and a larger interaction matrix element in this basis does not hold in general.
Here, one has to distinguish between the changes in the Wannier localization that are induced by a modification of the Wannier gauge from those caused by modifications of the lattice, caused e.g.\ by pressure.
The fact that for the discussed model, both methods to fix the Wannier gauge are equivalent, states that {\it for a given pressure}
the maximally localized Wannier functions yield the maximal Coulomb interaction matrix element.
Yet, as we shall see, a {\it pressure induced}  increase in the converged minimal Wannier spread is actually quite naturally concomitant with a  greater bare interaction.
\section{Results and discussion}
External pressure, or structural changes in general, provides an impetus to alter not only the one-particle band-structure of a material, but also the
Wannier functions.
Therefore, when investigating the pressure dependence of effective, i.e.\ screened, interaction matrix elements, one has to distinguish between influences of the former, which enter via a modification of screening channels, and of the latter that not only affects the polarization, but, on a more fundamental level, modifies already the bare Coulomb interaction matrix elements.
The fact that Wannier functions of a solid are not eigenfunctions of the system,
 may result in a nonstraightforward evolution when
parameters such as external pressure change.
\subsection{One-band tight-binding model in one dimension}
As a first model system, we investigate a tight-binding parametrization of a one-band solid in one dimension. 
In that case the maximally localized Wannier function is naturally given by the Fourier transform of the Bloch functions when inversion symmetry is verified~\cite{PhysRevB.26.4269,PhysRevB.56.12847}. 
Moreover, since no higher energy orbitals are present, no partial screening can occur [in \eref{Pr}~: $P_r=0$] and the Hubbard $U$
equals the on-site matrix element of the bare Coulomb interaction, $U=V_{\svek{R},\svek{R}}$.
\begin{figure}[t]
\begin{center}
\includegraphics*[angle=-90,width=.475\textwidth]{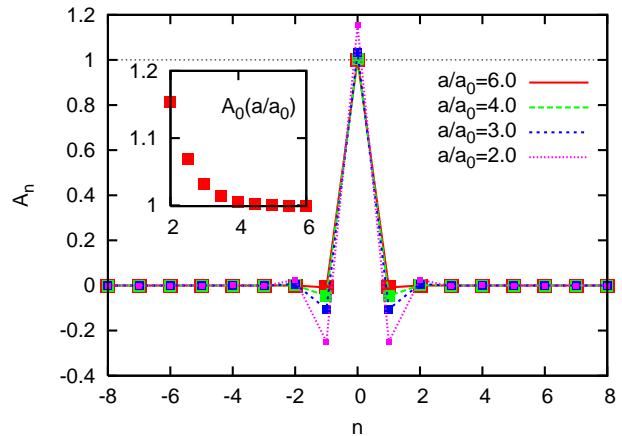}
			\caption{(Color online) The discrete distribution $A_n$ for different lattice constants, $a/a_0$, as a function of the atomic ``distance'', $n$, in real space.  The inset shows the dependence of $A_0$, i.e.\ the weight of the atomic function at the origin, on the lattice constant.}
\label{figA1}
\end{center}
\end{figure}
\subsubsection{Bloch function in tight binding}
As building blocks of the tight binding basis functions we opt for a hydrogen-like 1s orbital in one dimension
\begin{eqnarray}
	\chi(x)&=&\frac{1}{\sqrt{a_0}} e^{-\left|x\right|/a0}
	\label{1s}
\end{eqnarray}
with the Bohr radius $a_0$. 
This orbital is a solution to the time-independent Schr{\"o}dinger equation with a single binding delta potential~: It represents an eigenstate of the ``atomic'' Hamilton operator
\begin{equation}
H_0^{atom}=\frac{1}{2m_e}P^2-\frac{\hbar^2}{2m_ea_0}\delta(X)	
	\label{H0}
\end{equation}
with the eigenvalue $-\frac{\hbar^2}{2m_ea_0^2}$. While the tight-binding approach with this orbital can in principle be used to approximate the Bloch eigenfunction for a Hamiltonian with any potential, its use is obviously most justified for a Kronig-Penney type of model~\cite{KronigPenney,PhysRevB.44.5470} with a Dirac-comb potential.

The Bloch function is written as~:
\begin{eqnarray}
	\psi_k(x)&=&A_k\sum_R e^{\im kR}\chi(x-R)
\end{eqnarray}
Here, the factors $A_k$ assure the orthonormality of the Bloch function $\psi_k$, and is determined to be~: 
\begin{eqnarray}
	A_k&=& \left( 1+2\sum_{m=1,...,\infty}s_m\cos(km a)\right)^{-1/2}
\end{eqnarray}
where, with the lattice constant $a$, and the integer $m$, $m  a$ denotes the distance to the $m$th neighboring site. Further, $s_m$ denotes the overlap integral between the atomic function at the origin and its $m$th neighbor, and is given by~: 
\begin{eqnarray}
s_m&=&\int dx \chi^*(x)\chi(x-m a)\nonumber\\
&=&\left( 1+\frac{\left|m \right|a}{a_0} \right) e^{-{\left|m \right|}a/a0}
\label{overlap}
\end{eqnarray}
\subsubsection{Wannier function}
In one dimension, the maximally localized Wannier function, $\psi_R(x)$, is given by the usual Fourier transform of the Bloch function if inversion symmetry is verified~\cite{PhysRevB.26.4269,PhysRevB.56.12847}. Therewith
\begin{eqnarray}
	\psi_R(x)&=&\oo{C}\int_{-\pi/a}^{\pi/a} \frac{dk}{2\pi} e^{-\im kR}\psi_k(r)\nonumber\\
	&=&\oo{C}\sum_{R\pr}\int_{-\pi/a}^{\pi/a} \frac{dk}{2\pi} e^{-\im k(R-R\pr)} A_k \chi(x-R\pr)\label{Awann}\nonumber\\
	&=&\oo{C}\sum_{R\pr}A_{R-R\pr}\chi(x-R\pr)\\
	\psi_0(x)&=&\sum_n A_n \chi(x-na)\label{psiR} 
\end{eqnarray}
\begin{figure}[t]
\begin{center}
\includegraphics*[angle=-90,width=.475\textwidth]{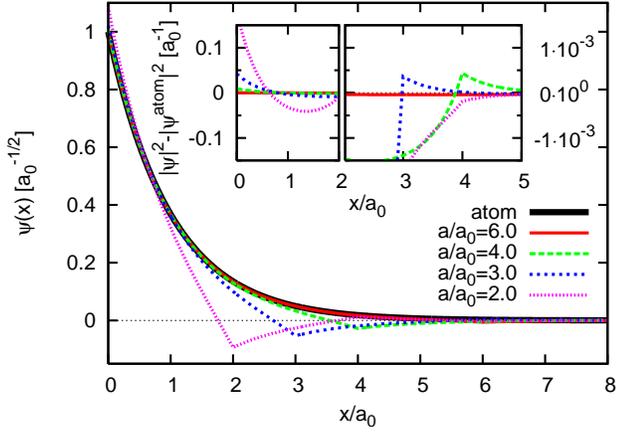}
			\caption{(Color online) Wannier functions for different lattice spacings $a/a_0$. The inset shows the deviation of the probability distribution $\left|\psi(x) \right|^2$ from the atomic limit (note the different scales in the left and right inner panel).}
\label{figpsi}
\end{center}
\end{figure}
where we defined
\begin{eqnarray}
A_{R-R\pr}&=&\int_{-\pi/a}^{\pi/a} \frac{dk}{2\pi} e^{-\im k(R-R\pr)}A_k	
\end{eqnarray}
or, with $R\pr=na$, and the reference $R=0$,
\begin{eqnarray}
A_n&=&\frac{\widetilde{A}_n}{C}=\oo{C}\int_{-\pi/a}^{\pi/a} \frac{dk}{2\pi} \frac{e^{\im kna}}{\sqrt{1+2\sum_{m =1}^\infty s_m\cos(km a)}}\nonumber\\
\label{An}
\end{eqnarray}
Demanding ${\int dx \left|\psi_R(x)\right|^2}=1$, the normalization constant $C$ becomes ~: 
\begin{eqnarray}
	C^2&=&\sum_n \sum_{m } \widetilde{A}_n \widetilde{A}_{n+m } s_m\label{AC}
	\label{Anorm}
\end{eqnarray}
The quantity $A_n$ is real for all $n$ for symmetry reasons. 
\fref{figA1} shows its behavior for different lattice constants~:
in the limit of large atomic separation ($a\gg a_0$), the overlaps $s_m$ are negligible, and the Wannier function $\psi_0(x)$, \eref{psiR},  will equal the atomic orbital $\chi(x)$, \eref{1s}. Thus the distribution $A_n$ picks up a single mode of the array, $A_n=\delta_{n,0}$ for the representative site ``0''. When pressure is applied, and the lattice constant shrinks, finite overlaps of the (non-orthogonal) hydrogen orbitals entails contributions from neighboring sites to mix in, and the 
distribution $A_n$ broadens (see \fref{figA1}). 
Since $A_{2n+1}\le 0$%
~\footnote{This can be seen (hand-wavingly) as follows. Consider only the first term in the $m $ sum. Then $A_n$ is defined for $s_1<0.5$, which is verified for $a/a_0>1.68$. For odd $n$, the relevant (=even) part of the periodic nominator, $\cos(kna)$ is always negative at the boundaries $\pm\pi$. On the other hand, $1/\sqrt{1+s_1\cos(ka)}$  is always positive ($s_1<0.5$), and becomes largest at the boundary. 
Thus the momentum integral that would vanish without the denominator, becomes negative, since the modulation of the denominator gives the negative values  a higher weight. Since $s_2\ll s_1$ the argument will remain valid when considering the complete sum over $m $.
 }%
, and $\left|A_{2\left|n\right|+1}\right|> A_{2(\left|n\right|+1)}$,
 the normalization of the Wannier function, \eref{Anorm}, causes the coefficient of the
atomic orbital at the origin to become larger than 1~: $A_0\ge 1$.
This results in a greater probability density $\left|\psi_{R}(x)\right|^2$ around the site origins, $x-R=0$, when pressure is applied.

The corresponding Wannier functions of the above cases are shown in \fref{figpsi}.
As anticipated from the above discussion, more weight is accumulated at the origin~: a harbinger for a larger on-site Coulomb interaction.
On the other hand, contrary to the atomic limit, the tails of the Wannier functions extend over several lattice constants, before the exponential decay sets in. This behavior at larger distance points to an increase in the Wannier spread and the growing of transfer integrals.
\begin{figure}[!b]
\begin{center}
\includegraphics*[angle=-90,width=.475\textwidth]{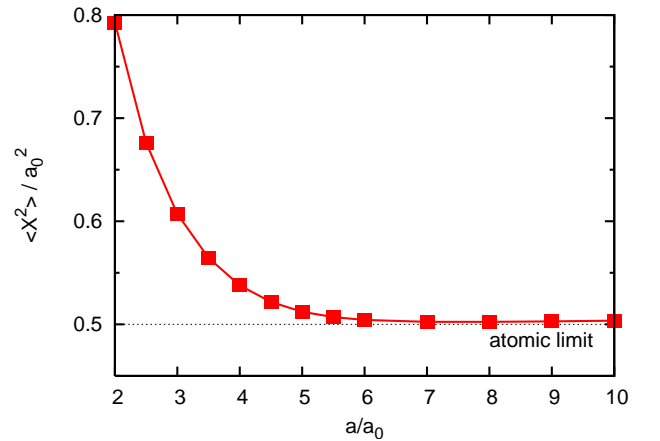}
			\caption{(Color online) Spread of the model as a function of lattice constant $a/a_0$. The dotted line indicates the atomic limit~: $\langle X^2\rangle=a_0^2/2$.}
\label{figA2}
\end{center}
\end{figure}
\subsubsection{Wannier spread}
Since by symmetry, $\langle r\rangle=\langle x\rangle=\bra{R=0}X\ket{R=0}= 0$,	
the spread of the Wannier function [$\Omega$ as given in \eref{spread}] reduces to~:
 \begin{eqnarray}
\left\langle X^2\right\rangle&=&	\bra{R=0}X^2\ket{R=0}=\int dx \,\,x^2\, \left|\psi_{R=0}(x)  \right|^2\nonumber\\
	&=&\sum_n \left|A_n\right|^2 \left(\frac{a_0^2}{2} +a^2n^2 \right)\nonumber\\
	&& +2\sum_n\sum_{m >0} A^*_nA_{n+m }    e^{-m a/a_0} \times\nonumber\\
	&&\times\left[ \left(\frac{m }{2}a\right)^3 \frac{2}{3a_0} +\frac{(m a)^2}{4} +\frac{m a}{2}a_0+\frac{a_0^2}{2} \right]             \nonumber\\
\end{eqnarray}
While the first term is always positive,  the second  will be negative for odd $m $ and positive for even neighbors (see the form of $A_n$).
Given the decay behavior of $A_n$, the second term will thus  be negative in total, yet small.
Indeed the major contribution to the spread comes from the first term, making it plausible that
the spread, as defined by $\langle X^2\rangle$, {\it increases} with a more widely distributed $A_n$, as shown in \fref{figA2}.

Although helpful for the understanding of the current model study, the spread is not too good a quantity for gaining quantitative insights from pressure studies of
realistic systems, as we will discuss later. 
\begin{figure}[b]
\begin{center}
\includegraphics*[angle=-90,width=.475\textwidth]{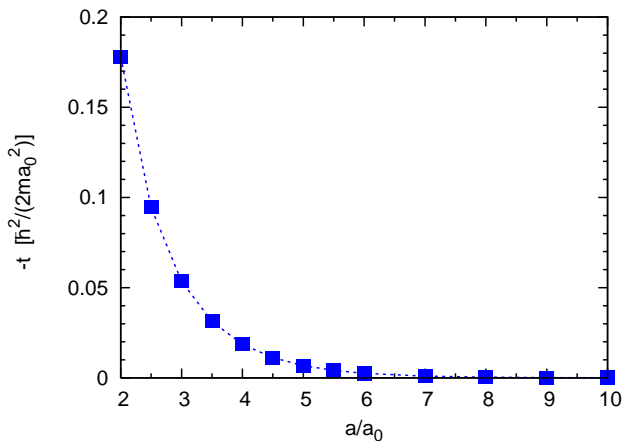}
			\caption{(Color online) Nearest neighbor transfer integral $t=\bra{\psi_{R=0}}H_0\ket{\psi_{R=a}}$ for a model with Dirac-comb potential, as a function of lattice constant $a/a_0$.}
\label{figAt}
\end{center}
\end{figure}
\subsubsection{Transfer integral}
For the transfer integral $t$, we need to explicitly specify the non-interacting Hamiltonian, $H_0$, and, for simplicity, we shall choose a Kronig-Penny-type model~\cite{KronigPenney,PhysRevB.44.5470} with an ionic Dirac-comb potential, i.e.\ 
\begin{eqnarray}
H_0&=&-\frac{1}{2m_e}P^2+\sum_lV^{ion}_l(X)	\nonumber\\
&=&-\frac{1}{2m_e}P^2-\sum_l\frac{\hbar^2}{m_ea_0}\delta(X-la) 
\end{eqnarray}
Then the nearest neighbor hopping $t=\bra{\psi_{R=0}}H_0\ket{\psi_{R=a}}$ can be expressed as~:
\begin{eqnarray}
	-\frac{2m_ea_0^2}{\hbar^2}\,	t &=&\sum_{m,n}A_nA_m  a_0s_{m+1-n} \\
	&+& \sum_{m,n}A_nA_m\sum_{l\ne m+1} e^{-\left|l-n\right|a/a_0}e^{-\left|l-m-1\right|a/a_0}\nonumber
\end{eqnarray}
As can be inferred from the dependence of the coefficients $A_n$, \eref{An}, and the overlap $s_m$, \eref{overlap},
the pressure induced delocalization increases the transfer integral, as expected.
\fref{figAt} displays the hopping as a function of the lattice constant.
\begin{figure}[b]
\begin{center}
\includegraphics*[angle=-90,width=.475\textwidth]{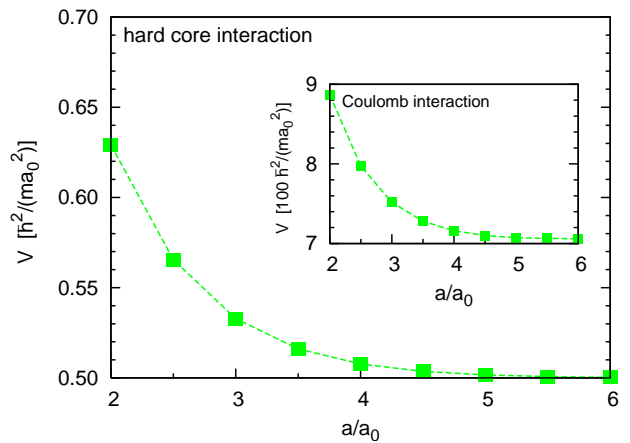}
			\caption{(Color online) On-site electron-electron interaction $V$, for a hardcore interaction (main graph), and a Coulombic one (inset), as a function of lattice constant $a/a_0$.}
\label{figA3}
\end{center}
\end{figure}
\subsubsection{Electron-electron interaction}
All of the above were concerned with the one-particle picture, i.e.\ the non-interacting Hamiltonian $H_0$. 
The on-site electron-electron interaction matrix element $V_{\svek{0},\svek{0}}$, that governs the two-particle term in the final many-body model, reads
 in the Wannier basis~:
\begin{eqnarray}
	V&=&V_{\svek{0},\svek{0}}\label{AV}=\int dx \int dx\pr \left|\psi_0(x)\right|^2 v(x,x\pr)\left|\psi_0(x\pr)\right|^2\\ [1ex]
	&=&\frac{\hbar^2}{m_ea_0}\left\{
	\begin{array}{lr}
	\displaystyle\int dx \left|\psi_0(x)\right|^4 &\hbox{(h.c.)}\\ [2ex]
	\displaystyle\int dx \left|\psi_0(x)\right|^2 \int dx\pr \frac{\left|\psi_0(x+x\pr)\right|^2}{\left|x\pr\right|} &\hbox{(C.)}\nonumber\\
	\end{array}\right.\nonumber
	%
\end{eqnarray}
when choosing a hardcore (h.c.) or Coulombic (C.) interaction~:
\begin{eqnarray}
	v(x,x\pr)&=&\frac{\hbar^2}{m_ea_0}\left\{
	\begin{array}{lc}
	\delta(x-x\pr) & \qquad\hbox{(hardcore)}\\ [2ex]
	\displaystyle\frac{1}{\left| x-x\pr\right|} &\qquad\hbox{(Coulomb)}\\
	\end{array}\right.
	\label{pot}
\end{eqnarray}
As shown in \fref{figA3} and anticipated before, the sharpening of the Wannier function at the origin causes  greater interaction matrix elements,
when pressure is applied, irrespective of the chosen electron-electron interaction.

As previously stated, the on-site electron-electron interaction $V$ from above equals the Hubbard $U$, since higher energy
orbitals, and thus screening effects, are absent by construction.
\subsubsection{Maximally localized Wannier functions versus maximal Hubbard interaction}
In Sec. I we mentioned that another technique to choose the Wannier function gauge is given by the request to maximize the static local, partially screened Coulomb interaction~\cite{RevModPhys.35.457,miyake:085122} -- the Hubbard $U$.

While not actually performing this approach for our model, we give evidence that in this simple case, both techniques are equivalent.
As said before, the Hubbard $U$ equals the bare Coulomb interaction $V$ since we consider only a single band, so, contrary to the general case~\cite{miyake:085122}, the argument does not involve any screening related effects.

The quest is thus to find a Wannier gauge, meaning an additional factor $\exp(\im\phi_k)$ in \eref{Awann}, that yields the greatest interaction element as given by \eref{AV}. The choice of gauge can be absorbed into the distribution $A_n$, and we define
\begin{eqnarray}
A_n[\phi]&=&\oo{C[\phi]}\int_{-\pi/a}^{\pi/a} \frac{dk}{2\pi} \frac{e^{\im kna} e^{\im\phi_k}}{\sqrt{1+2\sum_{m =1}^\infty s_m\cos(km a)}}\nonumber\\
\label{An2}
\end{eqnarray}
\begin{figure}[t]
\begin{center}
\includegraphics*[angle=-90,width=.475\textwidth]{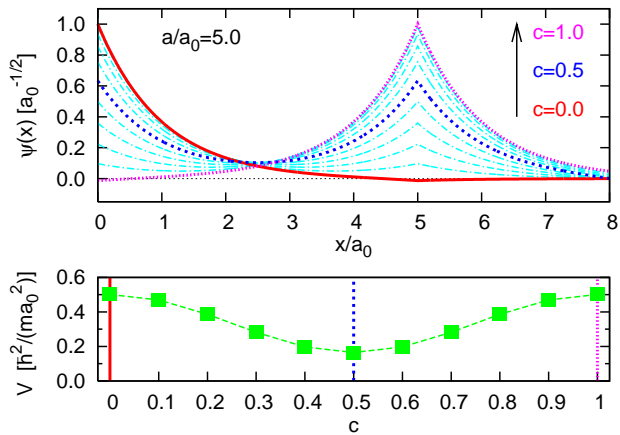}
			\caption{(Color online)  Wannier function $\psi_0(x)$ and corresponding on-site hardcore interaction (see \eref{AV}) as a function of the gauge parameter $c$ (see text for definition). A maximally localized Wannier function is obtained for $c=0.0$ (solid lines) and $c=1.0$ (dotted). 
			For $c=0.5$ (dashed), the Wannier function is symmetric with respect to $a/2$, and has equal weight at the site $R=0$ and $R=a$.
			The lattice constant is fixed at $a/a_0=5.0$. }
\label{figA4}
\end{center}
\end{figure}
Having seen the correspondence between the value of $A_0$ and the interaction strength, one might endeavor to solve the functional derivative $\delta A_0[\phi]/\delta\phi=0$ for $\phi$.
Yet the dependence of the normalization constant, $C$, given in \eref{AC}, makes this a rather tedious task analytically.
Instead, we shall argue for a specific example for the choice of Wannier gauge, and make some general comments.
Consider a gauge field that is linear in momentum, $\phi_k=-cka$ [see also Ref.~\cite{PhysRevB.44.5470}]. 
\fref{figA4} displays the Wannier function (upper panel) and the resulting on-site interaction (bottom panel) for the hardcore case (see \eref{pot}), for different gauge parameters $c$, and for a fixed lattice spacing $a/a_0=5.0$.
With $c\ne 0$ the inversion symmetry of the Wannier function with respect to the site $R=0$ is lost. This was the requirement for maximally localized Wannier functions in one dimension~\cite{PhysRevB.26.4269,PhysRevB.56.12847}, and as seen in \fref{figA4}, for $c>0$ the tail of the Wannier function augments, and the Coulomb interaction decreases.
While $c=1$ is a mere lattice translation, $c=\frac{1}{2}$ amounts to a case, where the Wannier function of site $R=0$ has equal positive weight
at $x=0$ and $x=a$, corresponding to $A_0=A_1$. 
As discussed above, an increase in $A_0$ is caused by the mixing in of negative components to $\psi_0$ from neighboring sites, leading to a smaller normalization factor $C$ in \eref{AC}. Owing to the symmetry, in the current example, $c=\frac{1}{2}$, negative contributions $A_m$ will come only from the sites $2m+1$ for $m<0$ and ${2m}$ for $m>0$. Yet the overall gain in renormalization is distributed over the two equivalent positions $m=0$, and  $1$ for which $A_m>0$.
As a result of this shifting of weight to the site $m=1$, $A_0$ and thus the on-site Coulomb interaction $V$ are much lower than in the case with inversion symmetry around that site. As seen in the bottom panel of \fref{figA4}, it is indeed the maximally localized Wannier functions ($c=0.0, 1.0$)
that yield the greatest possible Coulomb interaction matrix element for our simple model. 

For gauge fields $\phi_k=-ck^\alpha a$ depending on the momentum to the power $\alpha>1$, the argument is geometrically less obvious, but still true as verified numerically. Indeed, only the integrand in the coefficient $A_{n=0}$ of \eref{An2} is always positive for $\phi_k=0$ as a function of $k$. Therefore any modulation in
$\cos(\phi_k)$, with $\phi_k\ne0$, will decrease the corresponding integral to a greater extent than for $n\ne 0$, in which case the integrand  changes sign with $k$ already for $\phi_k=0$. 
As a consequence, the decrease in the $n=0$ contribution to the overall normalization $C$ 
will be greater on a relative scale than for $n\ne0$, and thus $A_0$ decreases with any non-constant $\phi_k\ne 0$.

This can be taken as a further indication that also for realistic systems, the maximally localized Wannier functions and the maximal Hubbard $U$ approach are generally giving the same results.
\subsection{fcc hydrogen under pressure}
\begin{figure}[t!]
  \begin{center}
\includegraphics[angle=-90,width=0.49\textwidth]{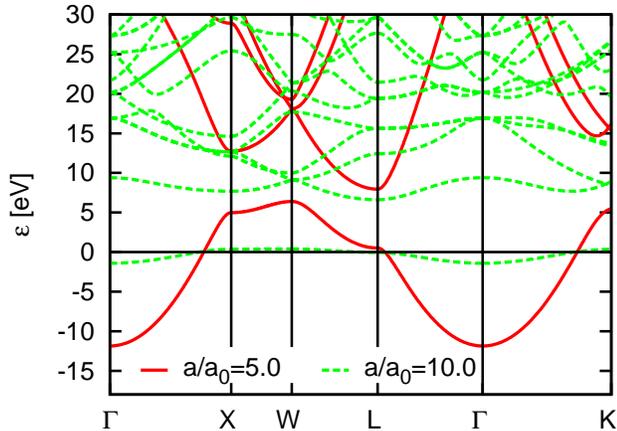}
    \caption{(Color online) LDA band-structure of fcc hydrogen for the two lattice spacings $a=5.0$~a.u. (solid) and $a=10.0$~a.u. (dashed). The Fermi level corresponds to the origin of energy.}
    \label{fccH_ek}
  \end{center}
\end{figure}
Towards a more realistic application of the gained insight, we apply the fully {\it ab initio} approach~\cite{miyake:085122} of the cRPA within maximally localized Wannier functions to the ``simplest'' realistic system, namely solid hydrogen. 

While at low pressure, solid hydrogen forms a molecular crystal. It was conjectured, already in the 1930s, that at high pressure
hydrogen should become an isotropic metal~\cite{wigner:764}. 
Here, however, we shall not be interested with the precise phase diagram of solid hydrogen. For the sake of simplicity, we 
assume throughout the discussion a face-centered cubic crystal structure with variable lattice constant.
Moreover, we are well aware that for the current case the problem of self-interaction~\cite{PhysRevB.23.5048} within the LDA formalism
is a particularly severe issue. However, here, we are not concerned with the accuracy of the LDA band-structure but with compression induced trends 
in a 3d multi-orbital setup.

For the one-particle band-structure, we employ in this work the LDA in the full-potential FP-LMTO~\cite{fplmto}
realization. 
In the LMTO basis, we include orbitals up to the 4f, using local orbitals~\cite{fplmto} for multiple orbitals per l-channel,  and use a Brillouin zone discretization with up to $10^3$ points.
As described in Ref.~\cite{miyake:085122}, maximally localized Wannier functions are then constructed for the hydrogen 1s band, which is entirely isolated from all other Kohn-Sham excitations. In other words, the effective model consists of a single half-filled orbital.
We stress that while the Wannier functions of different sites are orthogonal by constructions, the LMTO basis functions -- in analogy to the tight-binding parametrization in the preceding section -- are not. 
As a consequence, the same prototypical response to pressure as discussed above
 can be expected in the current case.

\fref{fccH_ek} shows the LDA band-structure for the extremal lattice constants that we consider. As expected, under growing compression, the dispersions increase and unoccupied bands are shifted upwards%
\footnote{The fact that the energy difference between the 1s and 2s,2p bands at $a=10.0$a.u. is already {\it smaller} than the atomic ionisation energy
is owing to the aforementioned spurious self-interaction in the LDA.
}%
.

\begin{figure}[b!h]
  \begin{center}
\includegraphics[angle=-90,width=0.475\textwidth]{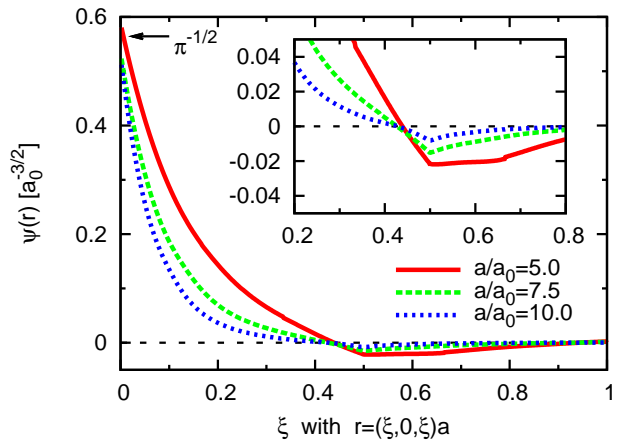}
    \caption{(Color online) Maximally localized Wannier function of the origin, as a function of the scaled distance $\xi$ towards the nearest neighbor atom at $r/a=(\xi=\frac{1}{2},0,\xi=\frac{1}{2})$. The arrow in the upper left indicates $1/\sqrt{\pi}$, which is the value of $\psi(0)$ for the exact atomic 1s hydrogen orbital (see text for comments). The inset shows a zoom on the position of the nearest neighbor atom. $10^3$ k points were used.}
    \label{fccHpsi}
  \end{center}
\end{figure}

\begin{figure}[t!h]
  \begin{center}
\includegraphics[angle=-90,width=0.4\textwidth]{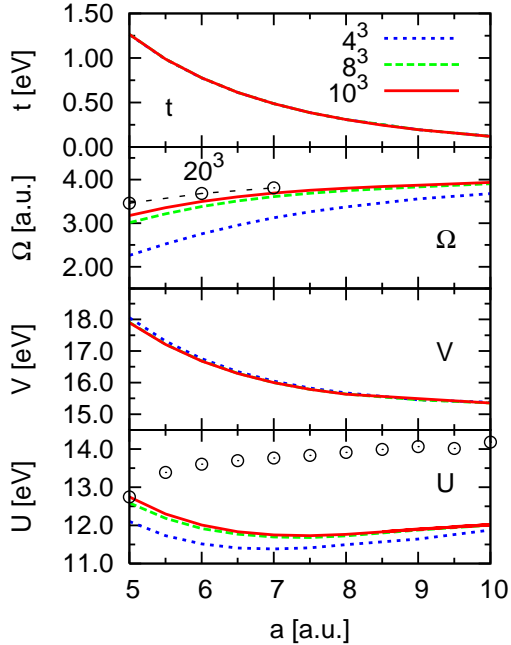}
    \caption{(Color online) Maximally localized Wannier functions and cRPA for the 1s orbital of fcc hydrogen. Shown are (from top to bottom) the nearest neighbor transfer integral (``hopping'') $t$, the spread $\Omega$ of the Wannier function (see \eref{spread}),
    the bare on-site Coulomb interaction $V_{\svek{R},\svek{R}}$,  and the cRPA screened on-site ``Hubbard'' interaction $U$
     as a function of the lattice constant $a$, and for different discretization of the Brillouin zone ($4^3$, $8^3$, $10^3$, and for the spread~: $20^3$)  as indicated. The symbols in the screened interaction $U$ indicate the values obtained when substituting in the cRPA polarization the LDA eigenvalues with those of the calculation for $a/a_0=5.0$ (using $10^3$ k-points).}
    \label{fccH}
  \end{center}
\end{figure}

In line with this is the behavior of the hopping $t$ --
the nearest neighbor transfer matrix element in basis of the maximally localized Wannier function for the subspace of the lowest Kohn-Sham  excitation. As shown in \fref{fccH}, it augments with decreasing lattice constant $a$,
accounting for the greater delocalization.
In \fref{fccHpsi} is shown the maximally localized Wannier function (it is real, cf. Ref.~\cite{PhysRevB.56.12847}) of the hydrogen atom at the origin,
as a function of the (scaled) distance $\xi$ towards the nearest neighbor at e.g.\ $r=(\xi=1,0,\xi=1)a/2$. 
When the lattice constant shrinks, clearly witnessed is both, a growing weight at the neighboring position ($\xi=1/2$), as well as an increase in the Wannier function at the origin ($\xi=0$). This is in complete analogy to what was shown in \fref{figpsi} for the model considered above\footnote{%
We note that the issue of self-interaction inherent in the LDA treatment impedes the Wannier function to converge towards the well-known 1s atomic eigenfunction $\psi(r)=1/\sqrt{\pi a_0^3}exp(-r/a_0)$. This is why the values of $\psi(0)$ for large separation are below teh indicated $1/\sqrt{\pi a_0^3}$. 
}%
.

It is thus expected that the on-site matrix element of the bare Coulomb interaction $V_{\svek{R},\svek{R}}$ grows under compression as before.
And, indeed this is the case, as can be inferred from \fref{fccH} in the second panel from the bottom.
Yet, does that entail for the Wannier spread $\Omega$ the same behavior as witnessed in the one dimensional model~?
In three dimensions, the gain in spread by effects of hybridizations with neighboring sites might turn out less prominent, since the region
occupied by nearest neighbor atoms is relatively small. Hence, the angular integral in $\Omega$, \eref{spread}, even for the radius of the distance to the 12 neighboring atoms (fcc) will run very much over a sphere on which the Wannier function is mostly zero.
And, indeed, as displayed in the second panel from the top in \fref{fccH}, the Wannier spread actually {\it decreases} under compression%
\footnote{We note that the {\it true} value of the spread at infinite separation should be $3a_0^2$ -- the atomic value. Again this is due to the
inadequateness of the LDA.
}%
.

A direct interpretation of the spread in pressure studies has thus to be taken with caution. First, depending on the crystal-structure (dimension, number and position of neighboring atoms) the spread does not necessarily reflect the increased delocalization of charge carrier
but can be dominated by the more isotropic concentration of weight at the origin (see \fref{fccHpsi}). Second, as seen in \fref{fccH} (and already in the seminal work, Ref.~\cite{PhysRevB.56.12847}, Table I), the momentum space convergence of $\Omega$ is poor.
Also, a change in lattice constant upon compression, changes the accuracy of the spread function when the number of k-points is kept constant. 
While the spread is of course the entity that is minimized in order to obtain the maximally localized Wannier functions for a given pressure, the spread itself is not a reliable measure for trends in the Wannier functions upon compression.
Instead, one should either plot the functions, or resort to the bare Coulomb interaction matrix elements, which -- owing to the different powers of the position operator -- converges well for a moderate k-point sampling.

In the current case, while still constructing an effective {\it one}-band model, the initial system contains higher energy orbitals.
As a result, and contrary to the simple tight-binding model, there is a non-zero partial polarization, $P_r$ of \eref{Pr}, that 
screens the bare Coulomb interaction $V$. The on-site part of the screened interaction ${W_r}_{\svek{0},\svek{0}}$ -- the Hubbard $U$ -- within cRPA is displayed
in the lowest panel of \fref{fccH}. 

Interestingly, this quantity, in contrast to the bare
interaction, shows a non-monotonic behavior, that reflects the struggle of two opposing effects in the polarization $P_r$.
As can be inferred from \eref{Pr}, and the equation for $P_s$, \eref{pol}, the changes in the polarization $P_r$ originate from modifications of transitions from the Wannier ``1s'' into orbitals at higher energy.
These can be altered by two ingredients (see again \eref{pol})~: the transition matrix elements, and the (Kohn-Sham) transition energies $\epsilon_{\svek{k}n}$, i.e.\ the band-structure.
The effect of pressure will be different for these two mechanisms. Indeed, the increasing compression pushes the bands further apart, as seen in \fref{fccH_ek}, {\it diminishing} the polarization. The increased overlaps/hybridizations of orbitals, on the other hand, tend to make the polarization {\it grow}.
In order to separate the influence of the two contributions, we computed the partially screened Coulomb interaction, $U$, as a function of lattice spacing, albeit while keeping the band-structure fixed at the values obtained for the highest pressure, $a/a_0=5.0$.
Thus the corresponding screened interaction will contain only the effect of changes in the Bloch functions.
As indicated by the symbols in the figure of the Hubbard U, the latter decreases under compression, proving the above conjectured opposition in trend to the influence of the band-structure.
 
This effect is surely very material specific. In systems with more electrons, where pressure will e.g.\ cause the enhancement of bonding/anti-bonding splittings, it can be expected that the changes in the band-structure often prevail such as to reduce the polarization under compression.
\section{Conclusions and Perspectives}
In conclusion, we have studied the influence of external pressure onto the construction of effective low energy many-body systems.
Using maximally localized Wannier functions for a one-band tight binding model, we rationalized the counter-intuitive, yet prototypical behavior of the bare Coulomb interaction, namely that its matrix elements augment upon compression, as a consequence of the delocalization of the Wannier functions.
This we understood to be caused by increased admixtures of non-orthogonal nearest neighbor tight-binding functions when the lattice spacing shrinks.
As a more realistic system, we investigated fcc hydrogen under pressure, and constructed an effective model for the half-filled 1s orbital using maximally localized Wannier functions. For the transfer integrals and the bare Coulomb interaction, we witnessed the same tendencies as in the model case. Yet, the Hubbard $U$, calculated as the on-site screened interaction within the constraint RPA technique, exhibited
a non-monotonous trend under compression.
This we traced back to a struggle between two opposing effects in the strength of screening.
All these highlight the intricacy of mechanisms that influence effective models, and emphasizes the need 
for reliable {\it ab initio} techniques for their construction.
\section*{Acknowledgments}
This work was in part supported by
the G-COE
program of MEXT(G-03) and Grant-in-Aid for Scientific Research from MEXT
Japan (Grants No. 19019013 and No. 19051016).


\end{document}